# Online aging study of high rate MRPC*

Jie Wang(王杰)[1,2;1], Yi Wang(王义)[1], S. Q. Feng(冯笙琴)[2], Bo Xie(谢波)[2], Pengfei Lv(吕鹏飞)[1],
Fuyue Wang(王扶月)[1], Baohong Guo(郭宝鸿)[1], Dong Han(韩冬)[1], Yuanjing Li(李元景)[1]

[1] Key Laboratory of Particle and Radiation Imaging, Tsinghua University, Beijing 100084, China

[2] College of Science, China Three Gorges University, Yichang 443002, China

**Abstract:** With the constant increase of accelerator luminosity, the rate requirements of the MRPC detectors become important. In the same time, aging problem of the detector has to be studied meticulously. An online aging test system is set up in our Lab. The setup of the system is described and the purpose is to study the performance stability during the long time running under high luminosity environment. The high rate MRPC has been irradiated by X-ray for 36 days and accumulated charge density reached $0.1C/cm^2$. No obvious performance degradation is observed for the detector.

**Key words:** high rate MRPC, aging test, current, counting rate

**PACS:** 29.40.Cs

## 1 Introduction

The multi-gap resistive plate chamber (MRPC) is a kind of new type of gas detector developed at the end of last century in European Organization for Nuclear Research [1, 2]. Due to the properties of excellent time resolution, high efficiency, low cost and easy production of large area, the MRPC is widely used in the construction of time of flight (TOF) system in particle physics and nuclear physics experiments such as the RHIC-STAR[3,4] and LHC-ALICE[5]. University of Science and Technology of China (USTC) and Tsinghua University designed MRPC prototype and constructed TOF system for RHIC-STAR and the TOF system plays an important role in STAR physics analysis such as the observation of antihelium-4 [6]. Common MRPC is assembled with float glass and its efficiency drops fast when the incident particle rate excess a few hundred $Hz/cm^2$. FAIR-CBM is a high luminosity experiment to study the QCD performance and equation of state and a MRPC based high rate TOF will be constructed [7]. In order to meet the rate requirement of CBM-TOF (~$25kHz/cm^2$ at center area), this kind of MRPC will be assembled with low resistive glass developed by Tsinghua University [8, 9]. As we know, the performance of MRPC will be degraded under irradiation and many works have been focused on this research [10-12]. This is the first time for the high rate MRPC used in large experiment and it is essential to test its aging performance. An online aging test system was established in our lab. The system consists of X-ray source, cosmic ray telescope, VME DAQ system and monitoring system. The high rate MRPC can be irradiated with X ray, in the same time, its efficiency and time resolution can be tested with cosmic ray system and the other performance such as working current and signal counting rate can also be recorded. This irradiation test is very similar to the detector test on GIF at CERN[13]. In this paper, a kind of high rate pad-readout MRPC has been irradiated by 60kV X-ray for 36 days and the accumulated charge on electrode reached $0.1C/cm^2$. No obvious performance degradation is observed for the MRPC.

## 2 Experimental apparatus and test environment

A pad readout MRPC aimed for the center area of CBM-TOF is tested. The structure of the module is shown in Fig.1. This double stacks module consists of ten gaps and the width of gap is 0.22 mm. There are 12 readout pads and the dimension of pad is 2cm ×2cm.

*Supported by National Natural Science Foundation of China (11420101004, 11461141011, 11275108). This study is also supported by the Ministry of Science and Technology under Grant no. 2015CB856905.

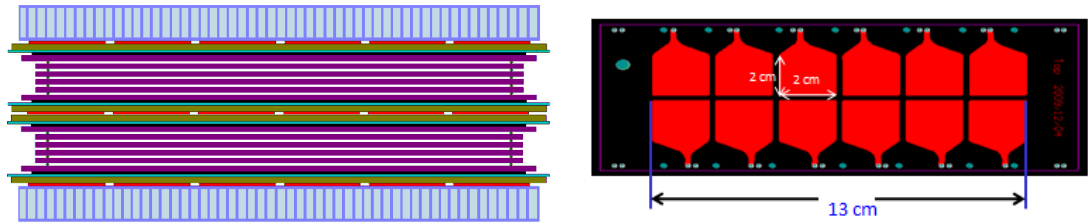

Fig.1 The structure of high rate MRPC module

In order to study how the detector performance change with irradiation dosage, the counting rate and working current are recorded online with Agilent 34410A Digital Multi-meter. Fig.2 shows the structure of the performance monitoring system. The MRPC signal from pad3 was sent to the meter which was recorded by a computer. The working current from the high-voltage source(CAEN N471)was connected to the other meter. The Spellman XRB80 X ray source is used to irradiate the detector and the signal rate and working current can be sampled every 30 seconds.

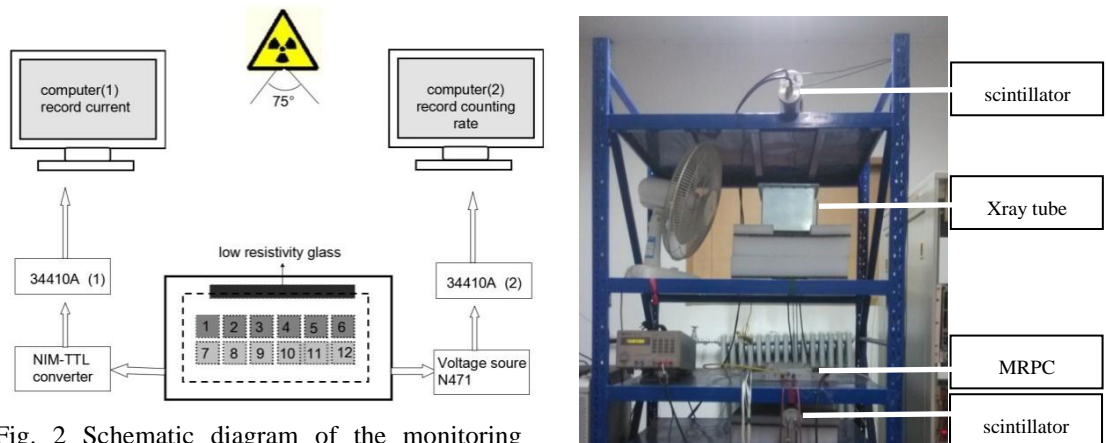

Fig. 2 Schematic diagram of the monitoring system

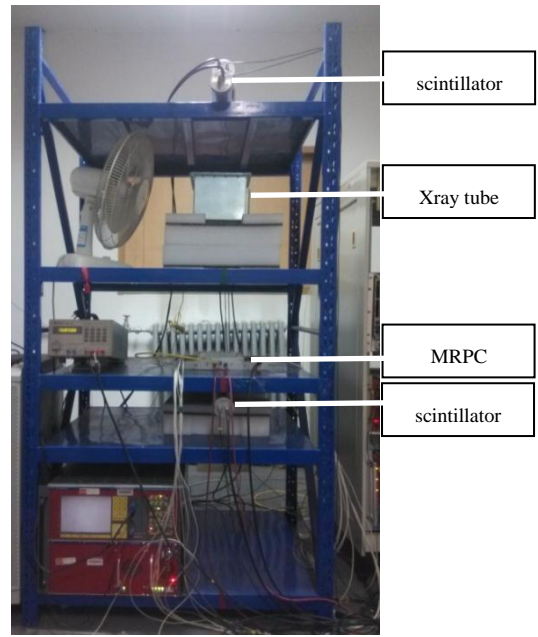

Fig. 3 The layout of the irradiation system

Experimental setup is shown in Fig.3. Two scintillator are 1.1 m apart and the distance from X ray source to MRPC is 60 cm. The HV of X ray source is set at 60 kV and the current at 0.4 mA. The irradiation is nearly uniform in each gap of MRPC by the simulation [14, 15]. The working gas consists of 90% freon, 5% iso-butane and 5% SF6 and the working voltage is ±6.2kV. The temperature in the room is kept at 24°C by air conditioner. A camera is put aside the gas mixture to monitor the flow rate of the working gas. During the experiment, the X-ray source has to be powered off for about 30 minutes every five days and we can check the noise rate and dark current in this interval.

## 3  Test results

The detector has been irradiated for 36 days. When the MRPC is irradiated by X-ray source, the current is around 1.4 μA and the signal counting rate is around 13 kHz/cm$^2$. In order to compare the performance, the efficiency is scanned before and after irradiation and the efficiency plateau is shown in Fig.4. It can be seen the efficiency becomes lower after irradiation when the working voltage is lower than 6kV. But at plateau, the efficiency is nearly the same.

We also check how the efficiency changes with the strength of X-ray. The results are shown in Fig.5. With the increase of current of the X ray source, a obvious decrease of efficiency of the MRPC is observed when the current increased to 0.8mA. So we can conclude that 0.8mA is the maxim irradiation strength. For the safety of MRPC, the working current of X ray source is set at 0.4mA and the irradiation is at half of the maxim strength.

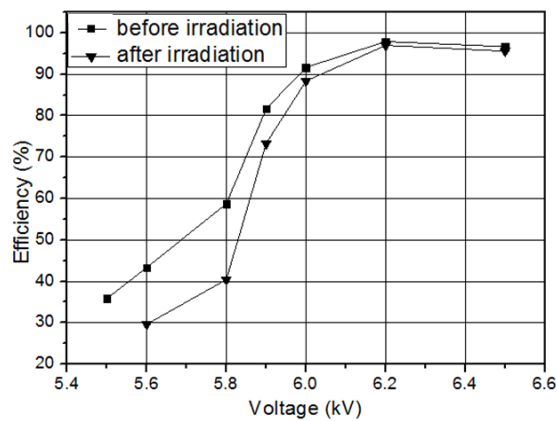

Fig.4 Efficiency as a function of high voltage before and after irradiation

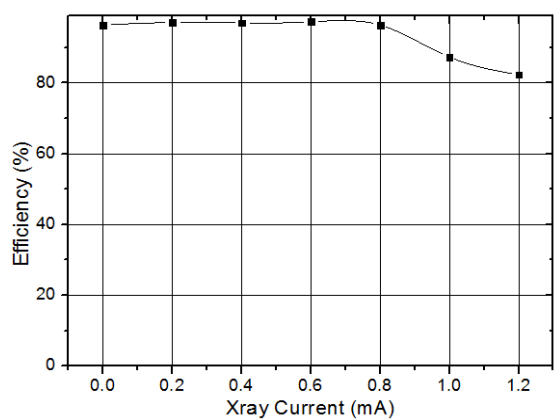

Fig.5 Efficiency vs the current of X-ray

Fig.6 shows the relation of working current and counting rate of the MRPC to the current of X-ray source. The current and counting rate of MRPC increases linearly with the source current. When the source current is 0.4 mA, the current of MRPC is 1.4μ A and the counting rate is 13k Hz/cm$^2$.

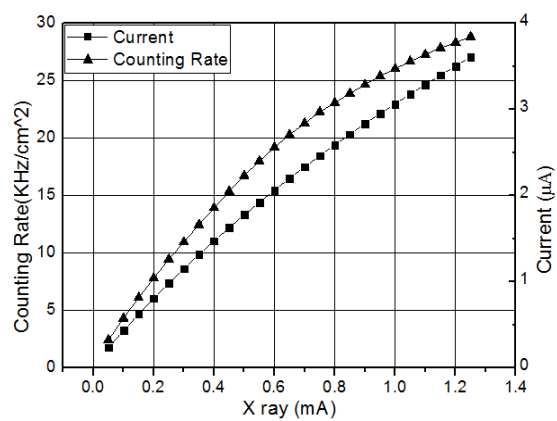

Fig.6 Current and counting rate change with x ray strength

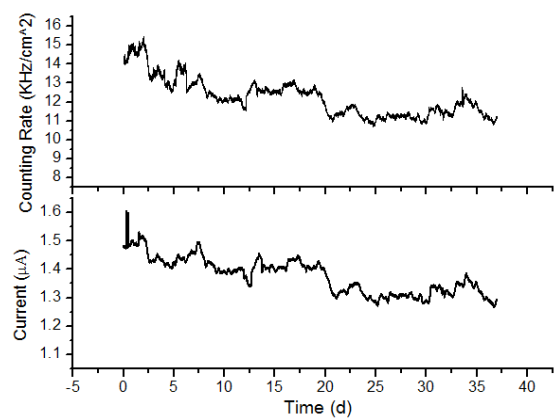

Fig.7 Current and counting rate change with irradiation time

To monitor the performance of MRPC, working current and counting rate are recorded per 30 seconds. Fig.7 shows the working current and counting rate change with irradiation time. It can be seen the working current and signal counting rate keep stable in the whole process. The little fluctuation may be caused by the X-ray strength fluctuation.

The efficiency, cluster size and noise are important parameters indicating the performance of MRPC. We can get from Fig.8 that these three parameters almost kept stable during the irradiation process. There is minor fluctuation. For example, the efficiency changes from 90% to 95% and the cluster size from 1.15 to 1.2. The noise kept at around 5 $Hz/cm^2$ before the irradiation and after a few hours of irradiation, the noise increased sharply to tens of $Hz/cm^2$ between $40Hz/cm^2$-$70Hz/cm^2$. But the noise gradually declines in the whole process.

Fig.9 shows time resolution measured during the whole irradiation process. Because of the statistic problem, the time resolution is analyzed by two methods. In the first method, it is analyzed every five days and only events (about 500) in the five days are used for analysis. The results are shown in Fig.9 (A). In the second method, time resolution is also analyzed every five days, but the events are all events we obtained from the first day. The results are shown in Fig.9 (B). It can be seen the fluctuation and error bars are all smaller in Fig.9 (B). The time performance keeps stable during the 36 days.

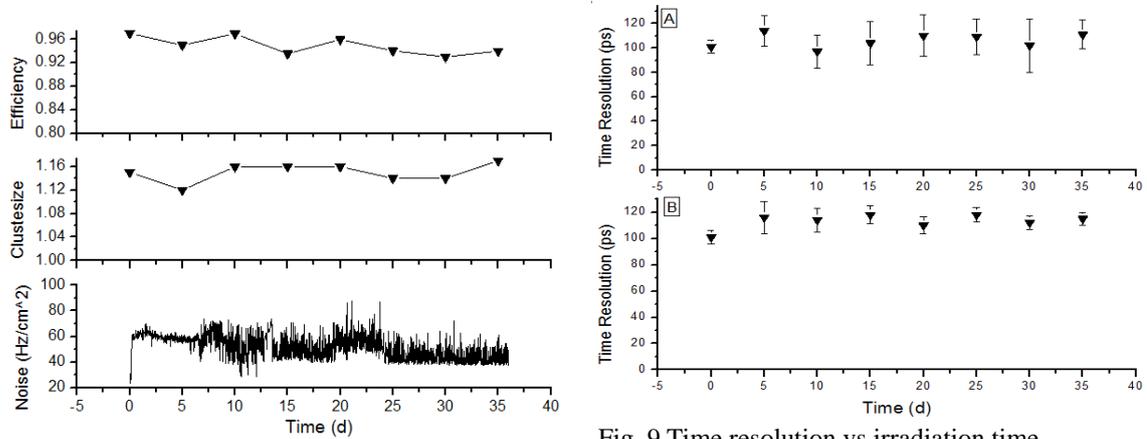

Fig.8 Efficiency, cluster size and noise change with irradiation time

Fig. 9 Time resolution vs irradiation time

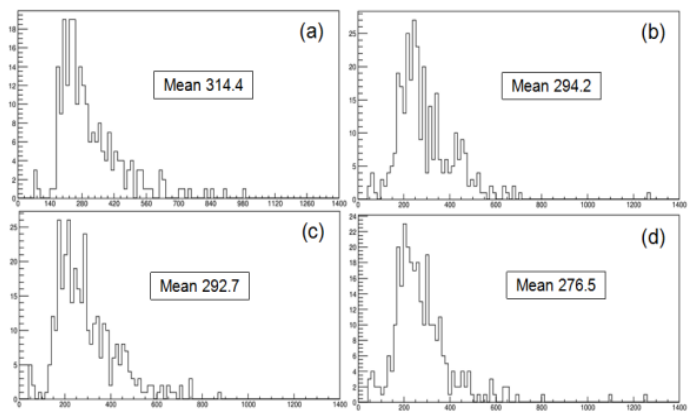

Fig. 10 The charge spectrum at different stage. (a) before irradiation, (b) after ten days irradiation, (c) after twenty days irradiation and (d) after thirty days irradiation.

The charge spectrum are shown in Fig.10 and it can be seen the average charge slightly decrease with time. This is caused mainly by space charge effect.

One phenomenon is observed that the performance recovery speed becomes slower after irradiation and these can be seen in Fig.11 and Fig.12. The black solid dots are experiments data and the red curves are fitted line. Fig.11 shows the current recovery after one day irradiation and Fig.11 the recovery after 36 days of irradiation. The current recovery curve can be fitted with exponential function such as:

$$C = a*\exp(-x/b) + C_0$$

C represents the current and x is time. a and $C_0$ are constants. The parameter b is 7.3 seconds and 13 hours in Fig.11 and Fig.12 respectively. The noise rate of the detector performs the same tendency as dark current. It can be seen the performance recovery speed is greatly affected by irradiation. This phenomenon has been observed at RHIC-STAR MTD system [16] and it is mainly caused by gas pollution effect. The gas pollution means the ionized gas is not exchanged in time which affects the MRPC performance.

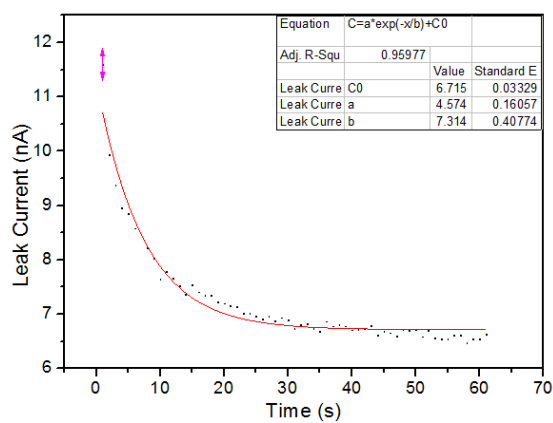

Fig.11 Current recovery curve after one day irradiation

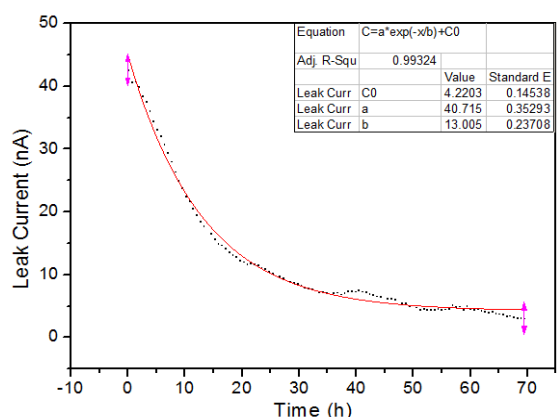

Fig.12 Current recovery curve after 36 days of irradiation

## 4 Conclusions

Application of high rate MRPC is very important in high luminosity hadron physics experiments. A kind of low resistive glass is developed in Tsinghua University and the rate capability of our high rate MRPC can reach 70 kHz/cm$^2$. In order to measure the long term stability of high rate MRPC, an online aging test system is set up in our Lab. The high rate MRPC has been irradiated by X-ray for 36 days and accumulated charge density reached 0.1C/cm$^2$. The working current and counting rate are monitored during the whole irradiation process. Other performance such as efficiency, time resolution and cluster size can be analyzed periodically. No obvious performance degradation is observed for working current, noise rate, time resolution and cluster size. But there is an obvious phenomenon that the dark current and noise recovery speed becomes slower after irradiation. Further analysis and measurements will be done later.